\documentstyle[sprocl,epsfig]{article}

\def\be{\begin{equation}}
\def\ee{\end{equation}}
\def\bea{\begin{eqnarray}}
\def\eea{\end{eqnarray}}
\newcommand{\case}[2]{\mbox{\footnotesize $\displaystyle \frac{#1}{#2}$}}
\newcommand{\lsim}{\mathrel{\rlap{\lower3pt\hbox{\hskip0pt$\sim$}}
\raise2pt\hbox{$<$}}}
\newcommand{\gsim}{\mathrel{\rlap{\lower3pt\hbox{\hskip0pt$\sim$}}
\raise2pt\hbox{$>$}}}

\begin{document}

\title{DYSON-SCHWINGER EQUATIONS \\ AND THE QUARK-GLUON PLASMA}

\author{C. D. ROBERTS AND S. SCHMIDT}

\address{Physics Division, Bldg. 203, Argonne National Laboratory\\
Argonne IL 60439-4843, USA}
% \\
% E-mail: cdroberts@anl.gov}


\maketitle\abstracts{We review applications of Dyson-Schwinger equations at
nonzero temperature, $T$, and chemical potential, $\mu$, touching topics such
as: deconfinement and chiral symmetry restoration; the behaviour of bulk
thermodynamic quantities; the $(T,\mu)$-dependence of hadron properties; and
the possibility of diquark condensation.}

\section{Introduction}
\label{intro}
Confinement and dynamical chiral symmetry breaking (DCSB) are consequences of
the little-understood long-range behaviour of the QCD interaction, and
developing a better understanding of that behaviour is a primary goal of
contemporary nuclear physics.  It is a prodigious problem whose solution
admits many complementary strategies.  Our approach is to apply a single
phenomenological framework to many observables, thereby identifying the
unifying qualitative features.  Non-hadronic electroweak interactions are the
best observables to study because the probes, the photon and $W$, $Z$ bosons,
are very well understood.  Following such applications$\,$\cite{ivanov} we
can infer consequences for QCD at extremes of temperature and chemical
potential.

Our tools of choice are the Dyson-Schwinger equations$\,$\cite{cdragw}
(DSEs), which at the simplest level provide a means of generating
perturbation theory and are an invaluable aid in proving renormalisability.
However, our interest stems from their essentially nonperturbative character.
For example, the DSE for the quark propagator is the QCD {\it gap equation}.
Its complete solution contains all that is necessary to describe DCSB and
yields insights into confinement, both of which are absent at any finite
order in perturbation theory.  Further, the Bethe-Salpeter equations (BSEs)
are just another form of DSE and these equations completely describe meson
structure.

The formulation of the DSEs is straightforward but their solution is not.
The equation for a particular propagator or vertex ($n$-point) function
involves at least one $m>n$-point function; e.g., the gap equation whose
solution is the dressed-quark propagator (2-point function) involves the
dressed-gluon propagator, a 2-point function, and the dressed-quark-gluon
vertex, a 3-point function.  Thus in the DSEs we have a countable infinity of
coupled equations and a tractable problem is only obtained if we truncate the
system.  This has been an impediment to their application: {\it a priori} it
can be difficult to judge the fidelity of a particular truncation scheme.
However, with expanding community involvement this barrier is being overcome
as truncation schemes are explored and efficacious ones developed.

\section{Gap equation}
\label{subsec:gapeqn}
The gap equation in QCD is the DSE for the quark propagator:
\begin{eqnarray}
\label{genS}
S_f(p)^{-1} & := & 
i \gamma\cdot p \,A_f(p^2) + B_f(p^2)  =
A_f(p^2) \left( i \gamma\cdot p + M_f(p^2) \right)\\
\label{gendse} & = & Z_2 (i\gamma\cdot p + m_f^{\rm bm})
+ Z_1\! \int^\Lambda_q 
g^2 D_{\mu\nu}(p-q) \frac{\lambda^a}{2}\gamma_\mu S_f(q)
\Gamma^{fa}_\nu(q,p),\\
\label{gluonprop}
D_{\mu\nu}(k) & = & \left(\delta_{\mu\nu} - \frac{k_\mu
                k_\nu}{k^2}\right)\frac{1}{k^2}\,{\cal P}(k^2)\,,
\end{eqnarray}
is the dressed-gluon propagator (in Landau gauge, just to be concrete),
$\Gamma^{fa}_\nu(q,p)$ is the dressed-quark-gluon vertex, $m_f^{\rm bm}$ is
the $\Lambda$-dependent bare $f$-quark current-mass and $\int^\Lambda_q :=
\int^\Lambda d^4 q/(2\pi)^4$ represents mnemonically a trans\-lationally-
invariant regularisation of the integral, with $\Lambda$ the regularisation
mass-scale.  The renormalisation constants for the quark-gluon-vertex, quark
wave function and mass: $Z_1(\zeta^2,\Lambda^2)$, $Z_2(\zeta^2,\Lambda^2)$
and $Z_m(\zeta^2,\Lambda^2) := Z_2(\zeta^2,\Lambda^2)^{-1}
Z_4(\zeta^2,\Lambda^2)$, depend on the renormalisation point, $\zeta$, and
the regularisation mass-scale.  (The renormalised current-quark mass is
$m_f(\zeta):=Z_m^{-1}m_f^{\rm bm}$.)

The qualitative features of the QCD solution of Eq.~(\ref{gendse}) are known.
The chiral limit is defined by $\hat m = 0$, where $\hat m$ is the
renormalisation-point-independent current-quark mass, and for
$p^2>20\,$GeV$^2$ the solution of Eq.~(\ref{gendse}) is$\,$\cite{mr97}
\begin{equation}
\label{Mchiral}
M_0(p^2) \stackrel{{\rm large}-p^2}{=}\,
\frac{2\pi^2\gamma_m}{3}\,\frac{\left(-\,\langle \bar q q \rangle^0\right)}
           {p^2
        \left(\case{1}{2}\ln\left[p^2/\Lambda_{\rm QCD}^2\right]
        \right)^{1-\gamma_m}}\,,
\end{equation}
where $\gamma_m=12/(33-2 N_f)$ is the gauge-independent mass anomalous
dimension and $\langle \bar q q \rangle^0$ is the
renormalisation-point-independent vacuum quark condensate.  The existence of
DCSB means that $\langle \bar q q \rangle^0 \neq 0$, however, its actual
value depends on the long-range behaviour of $D_{\mu\nu}(k)$ and
$\Gamma^{0a}_\nu(q,p)$, which is modelled in contemporary DSE studies.
Requiring a good description of light-meson observables necessitates $\langle
\bar q q \rangle^0\approx - (0.24\,\mbox{GeV})^3$.

The momentum-dependence in Eq.~(\ref{Mchiral}) is a crucial,
model-independent result because it is the {\it only} behaviour consistent
with the definition of the vacuum quark condensate as the trace of the
chiral-limit quark propagator:$\,$\cite{mr97}
\begin{equation}
-\langle \bar q q \rangle^0_\zeta= N_c\lim_{\Lambda\to\infty} \,
Z_4(\zeta^2,\Lambda^2)\, {\rm tr}_{D}\,\int_k^\Lambda\, S_0(k).
\end{equation}
Any model that generates
\begin{equation}
M_0(p^2) \sim p^{-2n}, \; n>1
\end{equation}
will yield $\langle \bar q q \rangle^0_\zeta \equiv 0$ from the definition of
the quark condensate.

Confinement is the absence of quark and gluon production thresholds in
colour-singlet-to-singlet ${\cal S}$-matrix amplitudes.  The absence of a
Lehmann representation for dressed-quark and -gluon propaga\-tors is
sufficient to ensure that.\cite{cdragw} Therefore the solution of
Eq.~(\ref{gendse}) can also yield information about confinement, as shown
clearly$\,$\cite{maris95} for QED$_3$.

Studies of Eq.~(\ref{gendse}) that employ a dressed-gluon propagator with a
strong infrared enhancement:$\,$\cite{pennington} ${\cal P}(k^2) \sim 1/k^2$,
and hence without a Lehmann representation, and $\Gamma^{fa}_\nu(q,p)$
regular in the infrared,\footnote{It is difficult to interpret particle-like
singularities in coloured Schwinger functions in a manner consistent with
confinement.$\,$\protect\cite{hawes}}
yield $S(p)$ that also does not have a Lehmann
representation.  Fine-tuning is not necessary.  Such models also easily
account for DCSB,$\,$\cite{mr97} with the correct value of $\langle \bar q q
\rangle^0$.

Contemporary DSE$\,$\cite{alkofer} and lattice$\,$\cite{cucchieri} studies
have reopened the possibility that ${\cal P}(k^2)\sim (k^2)^p$, $p\lsim 2$,
for $k^2 \simeq 0$; i.e., an infrared {\it suppression}.  The
phenomenological consequences of this have been re-explored:$\,$\cite{hawes}
when that ${\cal P}(k^2)$ obtained in contemporary lattice simulations is
used in Eq.~(\ref{gendse}) with an infrared-regular dressed-quark-gluon
vertex, DCSB does {\it not} occur and $S(p)$ has a Lehmann representation;
i.e, there is no signal of confinement.  The ${\cal P}(k^2)\sim (k^2)^p$-form
obtained in DSE studies can be made to support a nonzero condensate via
Eq.~(\ref{gendse}), however, its value is typically$\,$\cite{alkofer,jacques}
only 7-30\% of that required to explain observed phenomena, and again $S(p)$
does not exhibit signs of confinement.$\,$\cite{hawes}

\section{Exploring QCD at nonzero $T$ and $\mu$}
The dressed-quark propagator at nonzero-$(T,\mu)$ has the general form
\begin{eqnarray}
S(\tilde p_k) & = & \frac{1}{i\vec{\gamma}\cdot\vec{p}\, A(\tilde p_k) 
+ i\gamma_4 \omega_{k_+}\, C(\tilde p_k) 
+ B(\tilde p_k) }\,,\\
& = & -i \vec{\gamma}\cdot\vec{p} \, \sigma_A(\tilde p_k) 
- i\gamma_4 \omega_{k_+}\, \sigma_C(\tilde p_k) 
+ \sigma_B(\tilde p_k)\,,
\end{eqnarray}
where we have omitted the flavour label, $\tilde p_k =
(\vec{p},\omega_{k_+})$, $\omega_{k_+}= \omega_k + i \mu$ and $\omega_{k} =
(2k + 1)\pi T$, with $k\in {\rm Z}\!\!\!{\rm Z}$, is the quark's Matsubara
frequency.  The complex scalar functions: $A(\vec{p},\omega_{k_+})$,
$B(\vec{p},\omega_{k_+})$ and $C(\vec{p},\omega_{k_+})$ satisfy:
$ {\cal F}(\vec{p},\omega_{k_+})^\ast = {\cal F}(\vec{p},\omega_{-k_+-1})\,,
$
${\cal F}=A,B,C$, and although not explicitly indicated they are functions
only of $|\vec{p}|^2$ and $\omega_{k_+}^2$.  The dependence of these
functions on their arguments has important consequences in QCD: it can
provide an understanding of quark confinement and is the reason why bulk
thermodynamic quantities approach their ultrarelativistic limits slowly.  The
nonzero-$(T,\mu)$ gap equation is a straightforward
generalisation$\,$\cite{prl} of Eq.~(\ref{gendse}) and the Landau gauge
dressed-gluon propagator has the general form
\begin{eqnarray}
g^2 D_{\mu\nu}(\vec{p},\Omega_k) & = &
P_{\mu\nu}^L(\vec{p},\Omega_k) \,\Delta_F(\vec{p},\Omega_k) + 
P_{\mu\nu}^T(\vec{p})\, \Delta_G(\vec{p},\Omega_k) \,,\\
P_{\mu\nu}^T(\vec{p}) & := &\left\{
\begin{array}{ll}
0; \; & \mu\;{\rm and/or} \;\nu = 4,\\
\displaystyle
\delta_{ij} - \frac{p_i p_j}{|\vec{p}|^2}; \; & \mu,\nu=i,j=1,2,3
\end{array}\right.\,,
\end{eqnarray}
with $P_{\mu\nu}^T(p) + P_{\mu\nu}^L(p,p_4) = \delta_{\mu\nu}- p_\mu
p_\nu/({\sum_{\alpha=1}^4 \,p_\alpha p_\alpha})$; $\mu,\nu= 1,\ldots, 4$, and
$\Omega_k= 2 k \pi T$ the boson Matsubara frequency.

In studying the formation of a quark-gluon plasma, two transitions are
important: deconfinement and chiral symmetry restoration.  The simplest order
parameter for the chiral transition is
\begin{equation}
\label{chiorder}
{\cal X}(t,h) := {\sf Re}\,B_0(\vec{p}=0,\tilde \omega_0)\,;\;
t:= \frac{T}{T_c} - 1\,,\; h:= \frac{m^\zeta}{T}. 
\end{equation}
It is a general result that the zeroth Matsubara mode determines the
character of the chiral phase transition.

An order parameter for the deconfinement transition is realised$\,$\cite{prl}
via the Schwinger function:
\begin{equation}
\label{ft}
\Delta_{B_0}(x,\tau=0)
:=  T\sum_{n=-\infty}^\infty\,\frac{1}{2\pi^2 x}\int_0^\infty
dp \,p \,\sin(p x) \,\sigma_{B_0}(p,\omega_n),
\end{equation}
where we have set $\mu=0$ for illustrative simplicity.  If
$\sigma_{B_0}(p,\omega_n)$ has complex conjugate poles, $y_p$, then: 1) it
doesn't have a Lehmann representation; and 2) $\Delta_{B_0}(x,\tau=0)$ has
zeros.  The position of the first zero, $r_0^{z_1}(t)$, is inversely
proportional to Im$(y_p)$.  Thus
\begin{equation}
%\kappa_0(t):= \frac{1}{r_0^{z_1}(t)},
\kappa_0(t):= 1/r_0^{z_1}(t),
\end{equation}
is a confinement order parameter because $\kappa_0(t)\to 0$ as $t\to 0^-$
indicates that a temperature has been reached at which the poles have
migrated to the real-$p^2$ axis and the propagator has acquired a Lehmann
representation.  (This order parameter can be generalised to qualitatively
different functional realisations of the absence of a Lehmann
representation.)

\section{Locating the phase boundary in the $(T,\mu)$-plane}
DSE models constrained at $T=0=\mu$ can be used to estimate the location of
the phase boundary.  The studies we review all use rainbow truncation:
$\Gamma_\nu(q_{\omega_l};p_{\omega_k}) = \gamma_\nu$, which is the leading
term in a $1/N_c$-expansion of the vertex; and Landau gauge, with a
dressed-gluon propagator characterised by
\begin{eqnarray}
\label{uvpropf}
\Delta_F(p_{\Omega_k}) & = & {\cal D}(p_{\Omega_k};m_g)\,,\;
\Delta_G(p_{\Omega_k})  =  {\cal D}(p_{\Omega_k};0)\,,\\
\label{delta}
 {\cal D}(p_{\Omega_k};m_g) & := & 
        2\pi^2 D\,\case{2\pi}{T}\delta_{0\,k} \,\delta^3(\vec{p}) 
        + {\cal D}_{\rm M}(p_{\Omega_k};m_g)\,,
\end{eqnarray}
$p_{\Omega_k}=(\vec{p},\Omega_k)$, where $D$ is a mass-scale parameter and
${\cal D}_{\rm M}(p_{\Omega_k};m_g)$ may be large in the vicinity of
$p_{\Omega_k}^2=0$ but must be finite.

\subsection{$\mu=0$, $T_c=\,${\rm\bf ?}}
\label{subsec:D1}
The model obtained with $D= (8/9) m_t^2$ and
\begin{equation}
\label{modelfr}
{\cal D}_{\rm M}(p_{\Omega_k};m_g) = \case{16}{9}\pi^2\,
\frac{1-{\rm e}^{- s_{\Omega_k} /(4m_t^2)}}
        {s_{\Omega_k} }\,,
\end{equation} 
where $s_{\Omega_k}:= p_{\Omega_k}^2+ m_g^2$ [$m_g^2= 8\, \pi^2 T^2$ is a
gauge boson Debye mass], yields a finite-$T$ extension of a
phenomenologically efficacious one-parameter model dressed-gluon
propagator.$\,$\cite{prl} The mass-scale $m_t=0.69\,{\rm GeV}=1/0.29\,{\rm
fm}$ was fixed by requiring a good description of $\pi$- and $\rho$-meson
properties at $T=0$.  At a renormalisation point of $\zeta=9.47\,$GeV,
$m_u(\zeta)=1.1\,$MeV yields $m_\pi=140\,$MeV.

\begin{figure}[h,t]
\centering{\ \epsfig{figure=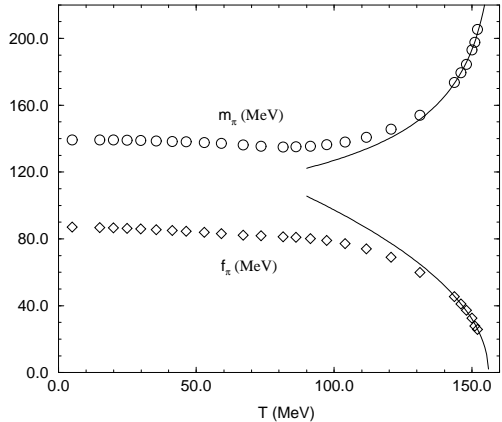,height=5.5cm}  }
\caption{\label{mpifpi} The pion mass and decay constant are independent of
temperature for $T\lsim 0.7 T_c^\chi$.}
\end{figure}

This model has {\it coincident} chiral symmetry restoration and deconfinement
transitions at
\begin{equation}
T_c^\chi = 0.15\,{\rm GeV}\, = T_c^{\kappa_0}
\end{equation}
with mean field critical exponents.  Studies that employ the rainbow
truncation {\it must} give mean field critical exponents$\,$\cite{arneb}
because contributions to the gap equation that describe the effects of
mesonic correlations, which are expected to dominate near the transition
temperature, can only arise as corrections to the vertex.  The behaviour of
$m_\pi$ and $f_\pi$ is depicted in Fig.~\ref{mpifpi}.

As a {\it bona fide} order parameter $f_\pi \propto (-t)^{1/2}$, which is
illustrated by the curve in Fig.~\ref{mpifpi}.  Hence, it follows from the
pseudoscalar mass formula:$\,$\cite{mr97} $f_\pi^2\,m_\pi^2 =
2\,m_u(\zeta)\langle\bar q q \rangle_\zeta^0$, that $m_\pi$ diverges at the
critical temperature; i.e., $m_\pi \propto (-t)^{-1/4}$, as illustrated.
Qualitatively, these two observations indicate that at $T=T_c^\chi$ there is
insufficient attraction in the pseudoscalar channel for a bound state to
form$\,$\cite{prl} and while correlations may persist above $T_c^\chi$ these
are properly identified as a continuum contribution to the pseudoscalar
vertex.

\subsection{$T=0$, $\mu_c=\,${\rm\bf ?}}
\label{subsec:D2}
The difficulties encountered in numerical simulations of lattice-QCD at
$\mu\neq 0$ are described in many contributions to this volume.  In studies
of the gap equation it only means that the self energies are complex-valued
functions.  The $T=0$ version of the model in the previous section is
obtained with
\begin{equation}
\frac{1}{k^2}\,{\cal P}(k^2):= 
\case{16}{9} \pi^2 \left[ 4 \pi^2 m_t^2 \delta^4(k)
+ \frac{1- {\rm e}^{-[k^2/(4 m_t^2)]}}{k^2}\right]
\end{equation}
in Eq.~(\ref{gluonprop}).  This model has$\,$\cite{greg} coincident, first
order deconfinement and chiral symmetry restoring transitions at $\mu_c=
0.375\,{\rm GeV}$, as measured by the location of the zero in the
$\mu$-dependent ``bag constant'':$\,$\cite{reg85} ${\cal B}(\mu)$.  It is
positive$\,$\footnote{The calculated value of ${\cal B}(0)=(0.104\,{\rm
GeV})^4=15\,{\rm MeV}/{\rm fm}^3$ is similar to that employed in bag-like
models.} when the Nambu-Goldstone phase is dynamically favoured; i.e., has
the highest pressure, and becomes negative when the Wigner pressure becomes
larger, which is why $\mu_c$ is the zero of ${\cal B}(\mu)$.  To gauge the
magnitude of $\mu_c$ we note that in a two-flavour free-quark gas the baryon
number density $\rho_B= 2 \mu^3/(3 \pi^2)$ so
\begin{equation}
\mu_c= 0.375\,{\rm GeV} \,\Rightarrow\, \rho_B^{u_F+d_F}= 2.9\,\rho_0,
\end{equation}
where $\rho_0=0.16\,{\rm fm}^{-3}$.  This may be compared with the central
core density of a $1.4 M_\odot$ neutron star: $3.6$-$4.1 \rho_0$, while $0.7
\mu_c$ corresponds to $\rho_0$.

In this model $m_\pi$ {\it decreases} slowly as $\mu$ increases, with
$m_\pi(0.7\,\mu_c)/m_\pi(0) \approx 0.94$.  At this point $m_\pi$ begins to
increase although, for $\mu<\mu_c$, $m_\pi(\mu)$ does not exceed $m_\pi(0)$,
which precludes pion condensation.  The behaviour of $m_\pi$ results from
mutually compensating increases in $f_\pi^2$ and $m(\zeta) \langle \bar q
q\rangle_\zeta^\pi$.  $f_\pi$ is insensitive to $\mu$ until $\mu\approx
0.7\,\mu_c$, when it increases sharply so that $f_\pi(\mu_c^-)/f_\pi(\mu=0)
\approx 1.25$.  At $\mu_c$, $m_\pi$ and $f_\pi$ drop discontinuously to
zero. The relative insensitivity of $m_\pi$ and $f_\pi$ to changes in $\mu$,
until very near $\mu_c$, mirrors the behaviour of these observables at
finite-$T$.$\,$\cite{prl} This study reveals an anticorrelation between the
$\mu$-dependence of $f_\pi$ and that of $m_\pi$.

\subsection{$T\neq 0$, $\mu\neq 0$}
\label{mnmodel}
This is a difficult problem and the most complete studies to
date$\,$\cite{thermo,basti} employ the simple {\it Ansatz} for the
dressed-gluon propagator obtained with $D= \eta^2/2$ and ${\cal D}_{\rm
M}(p_{\Omega_k};m_g) \equiv 0$ in Eq.~(\ref{delta}), and the mass-scale
$\eta=1.06\,$GeV fixed~\cite{mn83} by fitting $\pi$- and $\rho$-meson masses
at $T=0$.  With this {\it Ansatz} the gap equation is
\begin{equation}
\label{mndse}
S^{-1}(\vec{p},\omega_k) = S_0^{-1}(\vec{p},\tilde \omega_k)
        + \case{1}{4}\eta^2\gamma_\nu S(\vec{p},\tilde \omega_k) \gamma_\nu\,;
\end{equation}
and an integral equation is reduced to an algebraic equation whose solution
exhibits many of the qualitative features of more sophisticated models.

In the chiral limit Eq.~(\ref{mndse}) reduces to a quadratic equation for
$B(\tilde p_k)$, which has two qualitatively distinct solutions.  The
Nambu-Goldstone solution, with
\begin{eqnarray}
\label{ngsoln}
B(\tilde p_k) & = &\left\{
\begin{array}{lcl}
\sqrt{\eta^2 - 4 \tilde p_k^2}\,, & &{\sf Re}(\tilde p_k^2)<\case{\eta^2}{4}\\
0\,, & & {\rm otherwise}
\end{array}\,,\right.\\
C(\tilde p_k) & = &\left\{
\begin{array}{lcl}
2\,, & & {\sf Re}(\tilde p_k^2)<\case{\eta^2}{4}\\
\case{1}{2}\left( 1 + \sqrt{1 + \case{2 \eta^2}{\tilde p_k^2}}\right)
\,,& & {\rm otherwise}
\end{array}\,,\right.
\end{eqnarray}
describes a phase of this model in which: 1) chiral symmetry is dynamically
broken, because one has a nonzero quark mass-function, $B(\tilde p_k)$, in
the absence of a current-quark mass; and 2) the dressed-quarks are confined,
because the propagator described by these functions does not have a Lehmann
representation.  The alternative Wigner solution, for which
\begin{eqnarray}
\label{wsoln}
\hat B(\tilde p_k)  \equiv  0 &,\;& 
\hat C(\tilde p_k)  = 
\case{1}{2}\left( 1 + \sqrt{1 + \case{2 \eta^2}{\tilde p_k^2}}\right)\,,
\end{eqnarray}
describes a phase of the model with neither DCSB nor confinement.  

Here the relative stability of the competing phases is measured by a
$(T,\mu)$-dependent bag constant:$\,$\cite{thermo} ${\cal B}(T,\mu)$.  The
line ${\cal B}(T,\mu)=0$ defines the phase boundary, and the deconfinement
and chiral symmetry restoration transitions are coincident.  For $\mu=0$ the
transition is second order and the critical temperature is $T_c^0 =
0.159\,\eta = 0.17\,$GeV, just 12\% larger than the value reported in
Sec.~\ref{subsec:D1}.  For any $\mu \neq 0$ the transition is first-order and
the $T=0$ critical chemical potential is $\mu_c^0=0.3\,$GeV, \mbox{$\approx
30$\%} smaller than the result in Sec.~\ref{subsec:D2}.

The quark pressure, $P_q$, is easily calculated and $P_q \equiv 0$ in the
confined domain.$\,$\cite{thermo} However, this does {\it not} mean that the
vacuum is unaffected by changes in $(T,\mu)$.  On the contrary; e.g., in the
models described above, the condensate evolves with these changes, as it must
because it is a dynamical quantity.  At each $(T,\mu)$ the properties of the
hadronic excitations are calculated in the {\it evolved} vacuum and the
modification of the quark-constituents' propagation characteristics, which
the condensate's modification represents, makes a significant contribution to
the $(T,\mu)$-dependence of those properties.

In the deconfined domain, $P_q$ slowly approaches the ultrarelativistic, free
particle limit, $P_{\rm UR}$, at large values of $(T,\mu)$; e.g., at $ T \sim
0.3\,\eta \sim 2 T_c^0$, or $ \mu \sim \eta \sim 3 \mu_c^0$, $P_q \simeq
0.5\,P_{\rm UR}$.  This behaviour results from the persistence of momentum
dependent modifications of the quark propagator into the deconfined domain,
as evidenced by $C\not\equiv 1$ in Eq.~(\ref{wsoln}), which also entails a
``mirroring'' of finite-$T$ behaviour in the $\mu$-dependence of the bulk
thermodynamic quantities.

The $(T,\mu)$-dependence of vacuum and meson properties is easily calculated
in this model; e.g., the vacuum quark condensate is
\begin{equation}
\label{qbq}
-\langle \bar q q \rangle = 
\eta^3\,\frac{8 N_c}{\pi^2} \bar T\,\sum_{l=0}^{l_{\rm max}}\,
\int_0^{\bar\Lambda_l}\,dy\, y^2\,
{\sf Re}\left( \sqrt{\case{1}{4}- y^2 - \tilde\omega_{l}^2 }\right)\,,
\end{equation}
$\bar T=T/\eta$, $\bar \mu=\mu/\eta$; $l_{max}$ is the largest value of $l$
for which $\bar\omega^2_{l_{\rm max}}\leq (1/4)+\bar\mu^2$ and this also
specifies $\omega_{l_{max}}$, $\bar\Lambda^2 = \bar\omega^2_{l_{\rm
max}}-\bar\omega_l^2$, $\bar p_l = (\vec{y},\bar\omega_l+i\bar\mu)$.  At
$T=0=\mu$, $(-\langle \bar q q \rangle) = \eta^3 /(80\,\pi^2) = (0.11\,
\eta)^3$.  Obvious from Eq.~(\ref{qbq}) is that $(-\langle \bar q q \rangle)$
decreases continuously to zero with $T$ but {\it increases} with $\mu$, up to
$\mu_c(T)$ when it drops discontinuously to zero: as observed
elsewhere.$\,$\cite{prl,greg} That behaviour is a necessary consequence of
the momentum-dependence of the quark self energy, with the finite-$(T,\mu)$
behaviour of observables determined by
\begin{equation}
{\sf Re}(\omega_{[\mu]}^2)^d \sim [\pi^2 T^2 - \mu^2]^d\,,
\end{equation}
where $d$ is the observable's mass-dimension.  This is confirmed in the
chiral limit expression
\begin{eqnarray}
\label{npialg}
f_\pi^2 & = & \eta^2 \frac{16 N_c }{\pi^2} \bar T\,\sum_{l=0}^{l_{\rm max}}\,
\frac{\bar\Lambda_l^3}{3} \left( 1 + 4 \,\bar\mu^2 - 4 \,\bar\omega_l^2 -
\case{8}{5}\,\bar\Lambda_l^2 \right)\,.
\end{eqnarray}
The anticipated combination $\mu^2 - \omega_l^2$ appears and even without
calculation it is clear that $f_\pi$ will {\it decrease} with $T$ and {\it
increase} with $\mu$.

The $(T,\mu)$-response of meson masses is determined by the ladder BSE
\begin{equation}
\label{bse}
\Gamma_M(\tilde p_k;\check P_\ell)= - \frac{\eta^2}{4}\,
{\sf Re}\left\{\gamma_\mu\,
S(\tilde p_i +\case{1}{2} \check P_\ell)\,
\Gamma_M(\tilde p_i;\check P_\ell)\,
S(\tilde p_i -\case{1}{2} \check P_\ell)\,\gamma_\mu\right\}\,,
\end{equation}
where $\check P_\ell := (\vec{P},\Omega_\ell)$, with the bound state mass
obtained by considering $\check P_{\ell=0}$.  In this truncation the
$\omega$- and $\rho$-mesons are degenerate.

The pion solution of this equation is $\Gamma_\pi(P_0) = \gamma_5 (i \theta_1
+ \vec{\gamma}\cdot \vec{P} \,\theta_2 )$ and, consistent with what we saw
above, the mass is $(T,\mu)$-independent, until very near the transition
boundary.$\,$\cite{basti} For the $\rho$-meson the solution has two
components: one longitudinal, $\theta_{\rho+}$, and one transverse,
$\theta_{\rho-}$, to $\vec{P}$.  Equation~(\ref{bse}) yields an eigenvalue
equation for the bound state mass, $M_{\rho\pm}$, and using the chiral-limit
solutions, Eq.~(\ref{ngsoln}), one finds immediately that 
\begin{equation}
M_{\rho-}^2 = \eta^2/2,\;\mbox{{\it independent} of $T$ and $\mu$.}
\end{equation}
Even for $m\neq 0$, $M_{\rho-}$ changes by $< 1$\% as $(T,\mu)$ are increased
from zero toward their critical values.  This insensitivity is consistent
with the absence of a constant mass-shift in the transverse polarisation
tensor for a gauge-boson.

For the longitudinal component one obtains in the chiral limit:
\begin{equation}
\label{mplus}
M_{\rho+}^2 = \case{1}{2} \eta^2 - 4 (\mu^2 - \pi^2 T^2)\,.
\end{equation}
The combination $\mu^2 - \pi^2 T^2$ again indicates the anticorrelation
between the response of $M_{\rho+}$ to $T$ and its response to $\mu$, and,
like a gauge-boson Debye mass, that $M_{\rho+}^2$ rises linearly with $T^2$
for $\mu=0$.  The $m\neq 0$ solution for the longitudinal component is
semiquantitatively the same.

The BSE yields qualitatively the same behaviour for the $\phi$-meson.  The
transverse component is insensitive to $T$ and $\mu$, and the longitudinal
mass, $M_{\phi+}$, increases with $T$ and decreases with $\mu$.  Using $\eta
= 1.06\,$GeV, $M_{\phi\pm} = 1.02\,$GeV for $m_s = 180\,$MeV at $T=0=\mu$.

In a 2-flavour, free-quark gas at $T=0$ nuclear matter density corresponds to
$\mu= \mu_0 := 260\,$MeV$\,= 0.245\,\eta$ and the algebraic model yields
\begin{eqnarray}
\label{mrhoa}
M_{\rho+}(\mu_0)  \approx  0.75 M_{\rho+}(\mu=0) &,\; &
M_{\phi+}(\mu_0)  \approx  0.85 M_{\phi+}(\mu=0)\,.
\end{eqnarray}
Section~\ref{subsec:D2} indicates that a better representation of the
ultraviolet behaviour of $D_{\mu\nu}(k)$ increases the critical chemical
potential by 25\%.  This suggests that a more realistic estimate is obtained
by evaluating the mass at $\mu_0^\prime=0.20\,\eta$, which yields
\begin{eqnarray}
\label{mrhob}
M_{\rho+}(\mu_0^\prime) \approx  0.85 M_{\rho+}(\mu=0) &,\; &
M_{\phi+}(\mu_0^\prime) \approx  0.90 M_{\phi+}(\mu=0) \,;
\end{eqnarray}
a small, quantitative modification.  The difference between
Eqs.~(\ref{mrhoa}) and (\ref{mrhob}) is a measure of the theoretical
uncertainty in the estimates in each case.  Pursuing this suggestion further,
$\mu=\,^3\!\!\!\surd 2\,\mu_0^\prime$, corresponds to $2\rho_0$, at which
point $M_{\omega+}= M_{\rho+} \approx 0.72\, M_{\rho+}(\mu=0)$ and $M_{\phi+}
\approx 0.85\, M_{\phi+}(\mu=0)$, while at the $T=0$ critical chemical
potential, which corresponds to approximately $3\rho_0$ in
Sec.~\ref{subsec:D2}, $M_{\omega+}= M_{\rho+} \approx 0.65\,
M_{\rho+}(\mu=0)$ and $M_{\phi+} \approx 0.80\, M_{\phi+}(\mu=0)$.  These are
the maximum possible reductions in the meson masses.

\section{Diquark condensation}
A direct means of exploring the possibility that SU$(N_c)$ gauge theories
might support scalar diquark condensation is to study the gap equation
satisfied by
\begin{equation}
{\cal S}(p)^{-1} := 
\left(
\begin{array}{cc}
S(p)^{-1} & \Delta^i(p) \lambda_{\wedge}^i\, \tau^2_f\,\gamma_5  \\
 -\Delta^i(p) \lambda_{\wedge}^i\, \tau^2_f\,\gamma_5 
        & C (S(-p)^{-1})^{\rm T} C^\dagger
\end{array}\right),
\end{equation}
where $S(p)^{-1} = i \gamma\cdot p\, A(p^2) + B(p^2)$,
$\{\lambda_{\wedge}^i$, $i=1\ldots d_c$, $d_c= N_c (N_c-1)/2\}$ are the
antisymmetric generators of SU($N_c$), $C= \gamma_2\gamma_4$ is the charge
conjugation matrix, and here we consider SU($N_f=2$).  $\Delta^i(p) \neq 0$
for any $i$ indicates the formation of a diquark condensate.  ${\cal
S}(p)^{-1}$ is a matrix in the space of quark bispinors:
\begin{equation}
Q(x):= \case{1}{\surd 2}\left(\begin{array}{c}
                        q(x)\\
                        q_c(x) \end{array} \right),\;
\bar Q(x):= \case{1}{\surd 2}\left(\begin{array}{cc}
                        \bar q(x)\;
                        \bar q_c(x) \end{array} \right),
\end{equation}
$q_c:= -\bar q \,C$, $\bar q_c:= q^{\rm T} \,C$.  The gap equation is
\begin{eqnarray}
{\cal S}(p)^{-1} & = & {\cal S}_0(p)^{-1} 
+ \left( 
\begin{array}{cc}
\Sigma(p)_{11} & \Sigma(p)_{12}\\
\Sigma(p)_{21} & C \Sigma(-p)_{11}^{\rm T} C^\dagger \end{array} \right),
\end{eqnarray}
where ${\cal S}_0(p)^{-1} = {\rm diag}(i \gamma\cdot p + m, C (-i \gamma\cdot
p + m)^{\rm T} C^\dagger)$ and the form of $\Sigma(p)_{ij}$ specifies the
theory and its truncation.  This approach avoids a truncated bosonisation,
which in all but the simplest models is a procedure difficult to improve
systematically and prone to yielding misleading results.

\vspace*{0.5\baselineskip}

\hspace*{-\parindent}\underline{SU($N_c=2$)}:~Using the rainbow truncation and
a Feynman-like gauge for illustrative simplicity, the $T=0=\mu$ gap equation
in this theory yields
\begin{eqnarray}
p^2(A(p^2) - 1) & = & \case{3}{2} \,\int^\Lambda_k
g^2 D(p-k) \,p\cdot k \, \frac{A(k^2)}{d(k^2)},\\
\label{BD}
B(p^2) -m  & = &   \fbox{3}\,\int^\Lambda_k
g^2 D(p-k) \, \frac{B(k^2)}{d(k^2)},\\
\label{DD}
\Delta(p^2)  & = &   \fbox{3}\,\int^\Lambda_k
g^2 D(p-k) \, \frac{\Delta(k^2)}{d(k^2)},
\end{eqnarray}
where $d(p^2) = p^2 A(p^2)^2 + B(p^2)^2 + \Delta(p^2)^2$, and the
pseudo-reality of SU$(2)$ is responsible for the identical couplings in
Eqs.~(\ref{BD}) and (\ref{DD}).  Clearly, for $m=0$ the theory admits
degenerate and indistinguishable quark and (colour-singlet) diquark
condensates.  This result is valid independent of the truncation and gauge,
and is just one of the manifestations of Pauli-G\"ursey symmetry.  Similarly,
mesons and baryons, which are diquarks in SU$(N_c=2)$, are
degenerate.$\,$\cite{roberts} The phase structure of this theory at
nonzero-$(T,\mu)$ can certainly be very rich.

\vspace*{0.5\baselineskip}

\hspace*{-\parindent}\underline{SU($N_c=3$)}:~ In this case the interesting
possibility is the existence of a colour antitriplet diquark condensate:
$\{\lambda_\wedge^i\}_{i=1,2,3} = \{\lambda^2, \lambda^5, \lambda^7\}$.
Choosing the condensate to point in the $\lambda_\wedge^1$-direction, the
bispinor propagator separates into two pieces, one parallel and the other
perpendicular to the condensate's direction:
\begin{equation}
{\cal S}(p)^{-1} := 
\left(
\begin{array}{cc|cc}
S_{\|}(p)^{-1} I_2^c& 0 & \Delta^1(p) \lambda_{\wedge}^1\,
\tau^2_f\,\gamma_5 & 0\\
0 & S_{\perp}(p)^{-1} & 0 & 0 \\\hline
-\Delta^1(p) \lambda_{\wedge}^1\, \tau^2_f\,\gamma_5  & 0 & 
S_{\|}(p)^{-1} I_2^c & 0\\
0 & 0 & 0 & S_{\perp}(p)^{-1}
\end{array}\right).
\end{equation}
Here, since
\begin{equation}
\left(
\begin{array}{cc}
S_{\|}^{-1} I_2^c& 0 \\
0 & S_{\perp}^{-1} 
\end{array}\right) = 
I_3 \left(\case{2}{3} S_{\|}^{-1} + \case{1}{3}S_{\perp}^{-1}\right)
+ \case{1}{\surd 3}\lambda^8
\left( S_{\|}^{-1} -S_{\perp}^{-1}\right),
\end{equation}
the $ \lambda^a {\cal S} \lambda^a$ interaction in the gap equation {\it
couples} the parallel and perpendicular components.  In rainbow truncation
and using a Feynman-like gauge, the gap equation yields
\begin{eqnarray}
p^2(A_{\|}(p^2) - 1)  & = & \int^\Lambda_k 
g^2 D(p-k) \,p\cdot k \, 
\left[\frac{A_{\perp}(k^2)}{d_{\perp}(k^2)}
+ \case{5}{3}\frac{A_{\|}(k^2)}{d_{\|}(k^2)}\right],\\
p^2(A_{\perp}(p^2) - 1)  & = & \int^\Lambda_k 
g^2 D(p-k) \,p\cdot k \, 
\left[\case{2}{3}\frac{A_{\perp}(k^2)}{d_{\perp}(k^2)}
+ 2 \frac{A_{\|}(k^2)}{d_{\|}(k^2)}\right],\\
B_{\|}(p^2) - m  & = & \int^\Lambda_k 
g^2 D(p-k) \, 
\left[2 \frac{B_{\perp}(k^2)}{d_{\perp}(k^2)}
+ \case{10}{3}\frac{B_{\|}(k^2)}{d_{\|}(k^2)}\right],\\
B_{\perp}(p^2) - m  & = & \int^\Lambda_k 
g^2 D(p-k)  \, 
\left[\case{4}{3}\frac{B_{\perp}(k^2)}{d_{\perp}(k^2)}
+ 4 \frac{B_{\|}(k^2)}{d_{\|}(k^2)}\right],\\
\Delta^1(p^2)  & = &    \case{8}{3}\,\int^\Lambda_k 
g^2 D(p-k) \, \frac{\Delta^1(k^2)}{d_{\|}(k^2)},
\end{eqnarray}
$d_{\|}(p^2) = p^2 A_{\|}(p^2)^2 + B_{\|}(p^2)^2 + (\Delta^1(p^2))^2$ and
$d_{\perp}(p^2) = p^2 A_{\perp}(p^2)^2 + B_{\perp}(p^2)^2 $.

The class of models hitherto applied in exploring diquark
conden\-sation$\,$\cite{stephanov} can be characterised as those in which $
\int d\Omega_4\,p\cdot k\, D(p-k) = 0$.  In this class $A_{\|} = A_{\perp}
\equiv 1$ and when $B_{\|} = B_{\perp} \equiv 0$, which is always a solution,
$\Delta^1 \neq 0$ if the coupling is large enough.  Hence such restricted
models admit a rich phase structure at nonzero-$(T,\mu)$ because the $\|
\leftrightarrow \perp$ coupling is eliminated.

However, if one includes the next order contribution to the kernel of the gap
equation, the picture can change.  One study$\,$\cite{alkofernonR} suggests
that in that case, even without the $\| \leftrightarrow \perp$ coupling,
$\Delta^1 \equiv 0$ is the only solution at $T=0=\mu$.  The effect at
nonzero-$(T,\mu)$ of correcting the kernel has yet to be investigated but
this result signals a need for caution in making inferences about the phase
structure of QCD based on the rainbow-like truncation of this class of
models.

The more general class of models in which $\int d\Omega_4\,p\cdot k\, D(p-k)
\neq 0$ can be exemplified by the confining model introduced in
Sec.~\ref{mnmodel}.  In that case, if we consider $B_{\|}=B_{\perp}\equiv 0$,
the gap equation is solved with
\begin{equation}
\label{dqden}
p^2 A_{\|}(p^2)^2 + (\Delta^1(p^2))^2 = \case{1}{2}\eta^2
\end{equation}
and (setting $\eta^2 \to 1$) 
\begin{equation}
A_{\|}(p^2) = \case{1}{6}\left( 7 + 3 \sqrt{9 + 2/p^2}\right),\;
A_{\perp}(p^2) = \case{1}{2}\left( 3 + \sqrt{9 + 2/p^2}\right).
\end{equation}
However, inserting the result for $A_{\|}(p^2)$ into Eq.~(\ref{dqden}) yields
$(\Delta^1(p^2))^2 \leq 0$ for all $p^2$; i.e., $\Delta^1(p^2) \equiv 0$,
even in the rainbow truncation.  Thus diquark condensation at $T=0=\mu$ is
blocked by the $\| \leftrightarrow \perp$ coupling.  We expect that this
conclusion will be reinforced if the kernel is improved.$\,$\cite{brs} The
effect of $\mu\neq 0$ has not yet been explored but this result too advocates
caution in making inferences about the phase structure of QCD based on the
simple models hitherto employed.

\section{Epilogue}
Hadron observables are insensitive to the behaviour of the interaction at
$p^2 < \Lambda_{\rm QCD}^2$ and the rainbow truncation of the gap equation is
quantitatively reliable for $p^2 \gsim 1\,$GeV$^2$.  Thus the model
dependence in our approach is contained in an apparently small domain.
However, as illustrated by the material presented in this volume, even that
small domain of uncertainty admits a large variety of possibilities; although
apparently distinct {\it Ans\"atze} may really be different realisations of
the same phenomena.  It is also a crucial domain, covering that in which
mesonic correlations (vertex corrections) can influence quark propagation
characteristics, an effect that may become qualitatively important as the
phase boundary is approached.  Much has been achieved in localising the
model-dependence but more must be done to further ameliorate it.  Discussions
of the type represented by this volume are crucial to that endeavour.

\section*{Acknowledgments}
We are grateful to the staff of the ECT$^\ast$ for their assistance,
hospitality and support.  This work was supported by: the US Department of
Energy, Nuclear Physics Division, under contract no. W-31-109-ENG-38; the
National Science Foundation under grant no. INT-9603385; and S. Schmidt was
supported in part by the A.v.Humboldt-Foundation, as a Feodor-Lynen Fellow.

\begin{flushleft}

\end{flushleft}

\end{document}